\documentclass[12pt,a4paper,twoside]{article}
\usepackage[utf8]{inputenc}
\usepackage[english]{babel}
\usepackage[margin=3cm]{geometry}

\usepackage{hyperref}
\usepackage{graphicx}
\usepackage{times}
\usepackage{fancyhdr}
\usepackage{epsfig}
\usepackage{multirow}
\usepackage{amssymb,amsfonts,amsmath}
\usepackage{cite}

\usepackage{graphicx}
\newcommand{\mca}[1]{{\mathcal #1}}

\DeclareMathOperator{\rank}{rank}
\DeclareMathOperator{\tr}{tr}
\newcommand{\R}{\mathbb{R}}
\newcommand{\Rn}{\mathbb{R}^n}
\newcommand{\Rr}{\mathbb{R}^r}

\newcommand{\pan}[1]{\left\langle #1\right\rangle}
\newcommand{\de}[1]{\!\operatorname{d}\!{#1}}
\usepackage{color}

\newcommand{\mymodel}[2]{$\mathbf{#1.#2}$}
\newcommand{\mymodelm}[2]{\mathbf{#1.#2}}

\usepackage[hang,small,bf]{caption}

\begin{document}
	

\renewcommand{\multirowsetup}{\centering}

\thispagestyle{empty} 

\begin{center}
	{\Large\bfseries\sffamily
		Exact rank--reduction of network models\\
	}
	\vspace{1cm}\large
	{Eugenio Valdano\textsuperscript{1} and
		Alex Arenas\textsuperscript{2}}\\
	\vspace{1.1cm}
	{\footnotesize
		\textsuperscript{1}Center for Biomedical Modeling,  
		The Semel Institute for Neuroscience and Human Behavior,\\
		David Geffen School of Medicine,
		760 Westwood Plaza, \\
		University of California Los Angeles, 
		Los Angeles,
		CA 90024, USA.\\
		\textsuperscript{2}Departament d'Enginyeria Inform\`{a}tica i Matem\`{a}tiques,\\
		Universitat Rovira i Virgili, 43007 Tarragona, Spain.\\
	}
\end{center}

\vspace{2.2cm}

{\sffamily\footnotesize
	\centering{\bfseries Abstract}\\
	With the advent of the big data era, generative models of complex networks are becoming elusive from direct computational simulation. We present an exact, linear-algebraic reduction scheme of generative models of networks. By exploiting the bilinear structure of the matrix representation of the generative model, we separate its null eigenspace, and reduce the exact description of the generative model to a smaller vector space. After reduction, we group generative models in universality classes according to their rank and metric signature, and work out, in a computationally affordable way, their relevant properties (e.g., spectrum). The reduction also provides the environment for a simplified computation of their properties. The proposed scheme works for any generative model admitting a matrix representation, and will be very useful in the study of dynamical processes on networks, as well as in the understanding of generative models to come, according to the provided classification.
}

\newpage




\section{Introduction}

Network science is experiencing a burst of activity in the modeling and understanding of very large complex systems, including, for example, those formed by social interactions in microblogging platforms as Twitter~\cite{borge2016dynamics}, or by high-throughput molecular biology data related to genomes~\cite{halu2019}, proteomes~\cite{stelzl2005human}, metabolomes~\cite{brown2016metabolomics}, etc. However, this endeavor is limited by the computational effort required to simulate dynamical processes running on top of very large real networks. Moreover, the full knowledge of the connectivity structure (links) and dynamic state of their units (nodes) is often unaffordable. In this case the use of generative models of networks is the only alternative. 

Generative models of networks are a powerful tool for studying real-world networks and the dynamical processes unfolding on them. They can provide insight into network formation, by telling us which processes can (or cannot) lead to the development of certain descriptors. Their scope covers social networks~\cite{Granovetter1973}, neuroscience~\cite{Bassett2017}, human mobility~\cite{Barthelemy2011,Deville2016,Sole2016}, finance~\cite{Battiston2012}, ecology~\cite{Kefi2016a,Pilosof2017} and more. 
Through networked models, researchers have improved the understanding of complex systems in terms of easily interpretable analytical relations. A paramount example is the large literature on critical phenomena on complex networks: synchronization~\cite{arenas2008synchronization}, spreading processes~\cite{Newman2002,Guardi02,Barrat2008,PastorSatorras2015,bottcher2017critical,jgg2018,Skardal19}, or percolation~\cite{d2015anomalous,Hackett2016,Rapisardi2019}, to cite a few.
These models usually encode network structure into a small set of parameters and generative rules. In the last decades, along with the increased availability of highly-resolved data, generative models of networks have flourished to explain newly observed properties, like time-evolving contacts or multilayer topologies.
Generative models of networks have proven to be robust, accurate, and analytically treatable tools for describing families of networks, rather than single instances. We can say that they are still the tool of choice for uncovering mechanistic properties of complex systems that can be generalized to a wide set of contexts.

The main problem of this deluge of generative models is that, as they become richer and more intricate their dimensionality increases, and they become harder to simulate and analyze. 
Furthermore, the lack of an inclusive theoretical framework makes it difficult to derive theoretical relationships among models, which could tell us about how instances of different generative models are similar in their structure and functionality.

We propose an exact rank-reduction scheme for the matrix representation of generative models of networks (network models henceforth). We reduce the effective dimensionality of network models, facilitating their static and dynamical analysis. Moreover, this scheme allows us to define universality classes of network models in terms of the reduced features. We build a general, algorithmic derivation of some of the most relevant properties of these models, as their spectrum, and the behavior of some linear and nonlinear dynamical systems coupled to them.
We show that simple (low-rank) models explain the most important local properties present in real networks. We then define composition rules for models that allow us to reproduce more complex features, like mesoscale structures, while keeping the complexity of the models (in terms of number of variables necessary to reproduce the model) low.
Finally, we describe the relationship between models, and define equivalence classes of models, in terms of the action of symmetry groups on the rank-reduced spaces.
Our methodology provides a general framework for both classification and computation, that applies straightforwardly to future generative models, with no need to develop new ad-hoc approaches.

\section{Classification of network models}

The configuration model is one of the simplest and most popular generative network models~\cite{molloy1995critical}. One fixes the expected degree of each node, and considers all the network configurations (i.e., adjacency matrices) that respect the given degree sequence.
This generalizes to more complex models, which are always made up of a set of properties, and an ensemble of network configurations which is maximally entropic once the constraints induced by the properties hold.
The entry $A_{ij}$ of the matrix representation on such ensemble is proportional to the probability that the link $ij$ exists, and the properties defining the model completely determine the value of $A_{ij}$.
In the configuration model of $n$ nodes one has $A_{ij}=k_i k_j / \left( n \pan{k} \right)$, where $k_i$ is the expected degree of node $i$, and $\pan{k}$ is the expected average degree. 
We can write $A$ in matrix form: $A = KK^T$, with $K$ being the $n$-dimensional vector $K_i = k_i / \sqrt{n\pan{k}}$. Given that $A$ is the outer product of $K$ with itself, it is a rank-$1$ matrix: $\rank A = 1$, no matter the size of the system ($n$).

We argue that the rank of the matrix $A$ equals the number of node features the model constrains. The configuration model fixes only one feature per node --~the expected degree~-- resulting in $\rank A=1$. A model fixing two node features would result in $\rank A=2$, and so on. As a result, any symmetric, $n\times n$, matrix with $\rank A = r$ generates a model of a network of $n$ nodes which constrains $r$ properties per node. The general form of such a matrix is the linear combination of all the possible outer products among $r$ linearly independent vectors. We call them {\itshape metadegrees}, as they generalize the degree vector of the configuration model.
\begin{equation}
	A = \sum_{\mu, \nu=1}^r \Delta_{\mu\nu} v_\mu v_\nu^T.
	\label{eq:decomposition_indices}
\end{equation}
$v_\mu$ is an $n$-dimensional vector and represents the $\mu$-th metadegree, with $\mu=1,\cdots, r$. $\Delta$ is a $r\times r$ nonsingular matrix and encodes the coefficients of mixing among metadegrees.
The metadegrees can be arranged as columns of a $n\times r$ matrix $V$ whose entry $V_{i\mu}$ represents the value of the $\mu$-th metadegrees for the $i$-th node, leading to a pure matrix representation of $A$:
\begin{equation}
A = V \Delta V^T.
\label{eq:decomposition}
\end{equation}

Eq.~(\ref{eq:decomposition_indices},\ref{eq:decomposition}) hold for any value of $r=1,\cdots ,n$. However, we argue that $r$ must be much smaller than the size of the system ($r\ll n$), as useful physical models are usually designed to depend on few --fundamental-- parameters, compared to the size and complexity of the system under study. For this reason, we will use this decomposition to classify and easily solve large network models using their low-rank linear algebraic structure. Notwithstanding, we remark that models featuring nonlocal and mesoscale structures (like clustering, modularity, bipartiteness) apparently violate our statement, as they need high rank, some even a rank growing with the size of the system $n$. We will show that they can nonetheless be reduced to low-rank structures in Sec.~\ref{sec:promodels}.

  \begin{figure}\centering
\includegraphics[width=0.45\textwidth]{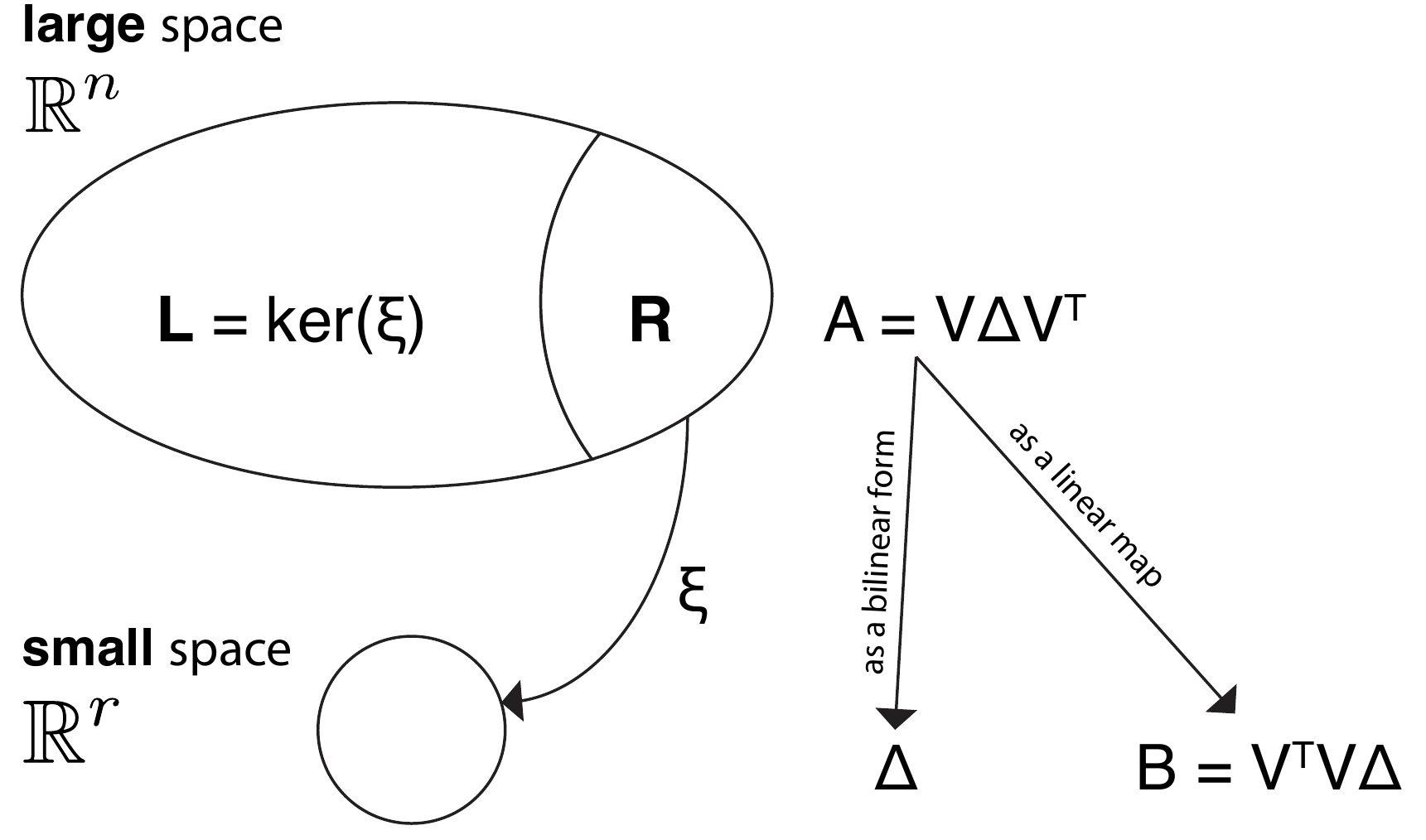}%
 \caption{\label{fig:schema}Representation of the exact rank-reduction scheme. The transformation laws for the matrix representation of the generative model are also shown. 
 }
 \end{figure}
%

The decomposition in Eq.~(\ref{eq:decomposition}) entails a powerful interpretation of the matrix representation of a generative model as a bilinear form $\Rn$. Given two vectors $X, Y \in \Rn$, the matrix representation returns their scalar product $X^T A Y$. This connection between the matrix representation and the structure of a scalar product on $\Rn$, already unveiled in~\cite{DeDomenico2013}, has deep implications in our study.

The main feature of our scalar product, however, is being highly degenerate. The {\itshape rank-nullity theorem} tells us that the eigenspace of $A$ associated to the eigenvalue $0$ has dimension $n-r$. This eigenspace, which we call $L$, is completely determined in terms of the kernel of the linear map $\xi : \Rn \rightarrow \Rr$, defined in terms of $V^T$: $\xi(x) = V^T x$. In other words $L$ is the set of vectors $x\in \Rn$ for which $V^T x = 0$. $L$ is the trivial subspace of $\Rn$. We focus on the restriction of $\xi$ to the subspace on which it is invertible. There, it induces an isomorphism between a small subspace $R$ of $\Rn$, with dimension $r$, and $\Rr$. 

Packing all this together, we induce a decomposition of the full space: $\Rn \simeq R \oplus L$, and this is the core step in our rank reduction. Given that an isomorphism is also a change of basis, $\Delta$ is the representation in $\Rr$ of the same scalar product that $A$ is in $R$. Moreover, given that by definition the restriction of $A$ to $L$ is identically zero, $\Delta$ encodes exactly the same information as $A$, once the trivial degeneracy is pruned. 
 
By Sylvester's law of inertia, we can always convert $\Delta$ to normal form, i.e. a diagonal matrix with entries $1$s and $-1$s (the number of positive entries of $\Delta$ is known as metric signature). Consequently, Eq.~(\ref{eq:decomposition}) is simply the scalar product $A$ written into normal form. A schematic representation of the exact reduction is presented in Fig.~\ref{fig:schema}.
 
 This decomposition has a straightforward consequence. It provides a universal classification of any possible matrix representation of a generative model for networks. All the possible network models (representable in terms of matrices) fall into classes defined by the rank $r$, and the metric signature $p$ of $\Delta$. We will denote these classes as (\mymodel{r}{p}).
Within each universality class, the values of metadegrees (columns of $V$) characterize a specific model, up to degeneracy induced by symmetry that we study in Sec.~\ref{sec:symmetry}.

 When $r=1$, the only existing class is (\mymodel{1}{1}) and it contains the configuration model. At $r=2$, we find two classes, which we can identify as {\itshape Euclidean} models (\mymodel{2}{2}) and {\itshape Lorentzian} models (\mymodel{2}{1}), according to the terminology used in general relativity.
 
 \subsection{Classification of popular generative models of networks}

Class \mymodel{2}{1} is particularly interesting, as it includes the {\itshape activity-driven model}~\cite{Perra2012}, a widely used model for time-evolving networks. This model assigns each node a probability of activation $a_i$. When active, a node establishes links with $m$ other random nodes (active or inactive). All links are reset before the next time step. The activity-driven model is an extremely simple model of temporal networks, yet, just as the configuration model, it has been successfully applied to many different contexts, and exhibits a rich and interesting macroscopic behavior. Given the absence of temporal correlations, it is fully represented by the matrix: $A = (m/n) (\Omega F^T + F \Omega^T )$, with $\Omega_i = a_i$ and $F_i = 1$, see~\cite{Valdano2018a}. From the previous expression, the rank-2 structure becomes apparent, as $A$ is the outer product of two linearly independent vectors. To show its signature, we have to write the metadegrees with $\Delta$ in normal form ($\Delta = \mbox{diag}(1, -1)$):
\begin{align}
v_1 &= \sqrt{\frac{m}{2n}} \left. \pan{a^2} \right.^{1/4} \left( F + \frac{\Omega}{\sqrt{\pan{a^2}}} \right)  ;\\
v_2 &=\sqrt{\frac{m}{2n}} \left. \pan{a^2} \right.^{1/4} \left(  F - \frac{\Omega}{\sqrt{\pan{a^2}}}  \right).
\label{eq:adm}
\end{align}
In a recent extension of the activity-driven model intended to mimic preferentiality in attachment~\cite{Alessandretti2017}, nodes are assigned specific values of attractiveness, in addition to the activity potential $a_i$. When a node activates, it will then be more likely to choose nodes with high attractiveness. This model falls again in class (\mymodel{2}{1}), and its reduced rank representation is the same as in Eq.~(\ref{eq:adm}), with a vector proportional to the attractiveness instead of the constant vector $F$ of the original activity model (see Appendix~\ref{sec:adm}). Also another extension, the simplicial activity-driven model~\cite{Petri2018}, where active nodes create cliques, instead of single links, to account for multi-agent interactions, can be accommodated in our classification. Depending on the relation between the average activity and the clique size, it can be easily proven that the simplicial activity-driven model is Euclidean (\mymodel{2}{2}) or Lorentzian (\mymodel{2}{1}). Specifically, one can prove that it is Euclidean if the average node activity $\pan{a}$ is higher than a threshold value: $\pan{a} > \left[ 2 (q-2) \right]^{-1}$, where $q$ is size of the clique. When clique size is not fixed, but follows a given distribution, the threshold value is more involute but can still be computed. Appendix~\ref{sec:adm} contains details of the computation regarding the activity-driven model and its generalizations.

The configuration model with degree-degree correlations also falls in the rank $r=2$ universality class. By setting the first metadegree vector of the model to be the degree vector $K$, one can make the Euclidean model \mymodel{2}{2} exhibit arbitrarily disassortative or assortative behavior~\cite{newman2002assortative} by tuning the second metadegree. Appendix~\ref{sec:ddcorr} reports the details of the calculations.

Finally, we can rank--reduce another popular generative model: the celebrated {\itshape stochastic blockmodel}~\cite{Holland1983}. It has a wide range of applications, as, for instance, community detection~\cite{Peixoto2017}.
In its simpler form, nodes are divided into $c$ subsets. Links within and between subsets occur with different probabilities. We define $k$ to be the average number of connections a node establishes with nodes from the same subset, and $h$ from subsets other than its own.
One can show that the rank of the resulting matrix representation is equal to the number of subsets: $r=c$. The signature exhibits instead two regimes. If subsets reflect a community structure, and nodes are more likely to connect within the same subset ($k>h/(c-1)$), then the model is Euclidean, and belongs to the class (\mymodel{c}{c}). If, instead, nodes tend to form link across subsets ($k<h/(c-1)$), the metric has Lorentzian signature, and the model belongs to the family class (\mymodel{c}{1}). Note that this particular classification of stochastic blockmodels separates modular structure from bipartite structures.
Even in the presence of a fine partition (large $c$), the only possible signatures are either Euclidean or Lorentzian. This will prove useful in what follows, when we study the internal symmetries of the different models. Appendix~\ref{sec:sbm} contains the details of the calculation.

\section{Derivation of model properties}

In addition to the classification, our rank-reduction scheme allows us to derive key properties of the models, from their general form (Eq.~(\ref{eq:decomposition})). Therefore, the results apply straightforwardly to all models --~already discovered or unknown~--, with no further need for ad-hoc approaches.
In what follows, we investigate the spectrum, as well as the solution of some linear and nonlinear processes. We start from the spectrum of the matrix representation of the model, which plays a key role in many centrality measures and determines the behavior of several dynamical processes. The largest eigenvalue \cite{lovasz1993random}, for instance, determines the critical behavior of synchronization~\cite{arenas2008synchronization} and diffusion~\cite{gomez2010discrete,Granell13,PastorSatorras2015,Valdano2015b} phenomena.
We wonder if our rank-reduction preserves the spectrum. We focus on the eigendecomposition of $A$ in the subspace $R$, which is itself the direct sum of the eigenspaces of $A$ relative to its nonzero eigenvectors. As mentioned, Eq.~(\ref{eq:decomposition}) is a change of basis for a bilinear form. Given that now we wish to preserve the spectrum, we need to treat $A$ as a linear map $\Rn\rightarrow\Rn$. The new representation in $\Rr$ is $B = \xi A \xi^{-1} = J\Delta$, with $J = V^T V$ (see also Fig.~\ref{fig:schema}). The explicit expression of the inverse isomorphism is $\xi^{-1} = V J^{-1}$.
Note that matrix $J$ encodes all the possible scalar products among the metadegrees: $J_{\mu\nu} = v_\mu \cdot v_\nu$.
We can give $J$ a useful statistical interpretation. Let us assume the values of metadegrees of each node ($v_{\mu, i} = V_{i\mu}$) come from a given probability distribution: the {\itshape metadegree distribution}. The metadegree distribution is the generalization of the degree distribution beyond rank $r=1$. Then, node metadegrees are samples from the metadegree distribution. As a result, the scalar product between $v_\mu$ and $v_\nu$ is proportional to the sample estimate of the expectation value of the product of these two metadegrees, intended as stochastic variables. In the limit of large network ($n \rightarrow \infty$) , the sample estimate converges to the true expected value: $(v_\mu \cdot v_\nu)/n \rightarrow \pan{v_\mu v_\nu}$.
This implies that the eigenvalues of the matrix representation of the generative model are linear combinations of the second moments of the metadegrees. This feature is extremely relevant, for instance, in the context of epidemic spreading. Many seminal works have shown the epidemic threshold of the configuration model to depend on the second moment of the degree~\cite{PastorSatorras2001,Newman2002}, with important implications for disease containment. More recently, the same property was found for the activity distribution and other models~\cite{Ferreri2014}. We now discover that this is a general property of any model, not a peculiarity of those two.

Through rank-reduction we can derive a simple formula for the eigenvectors, too. Let $\Lambda_\mu\not= 0$ an eigenvalue of $A$ (and $B$) and let $f^{(\mu)}$ be an associate eigenvector of $B$ in the reduced space ($Bf^{(\mu)} = \Lambda_\mu f^{(\mu)}$). Then $g^{(\mu)} = V\Delta f^{(\mu)}$ will be an eigenvector of $A$ for the same eigenvalue. The eigenvalues end eigenvectors of $B$ thus completely determine the spectral decomposition of $A$. Furthermore, one can choose $\{f^{(\mu)}\}$ to be a specific orthogonal basis of $\Rr$ with respect to the scalar product $\Delta$: $f^{(\mu)T}\Delta f^{(\nu)} = \delta_{\mu\nu} \Lambda_\mu^{-1}$. Then, $\{g^{(\mu)}\}$ will automatically be an orthonormal basis of $R$, thus completing an algorithmic construction of the spectral decomposition of the matrix from its rank-reduced transform.

In addition, the spectrum of $A$ completely determines the behavior of any linear diffusion process of the form $\dot{x} = (a+bA)x$, which can then be solved in the reduced space and then projected back. We can, however, use rank-reduction to solve a large class of nonlinear dynamical processes, too. Consider the following equation for the operator $X(t)\in \R^{n,n}$ (an $n\times n$ matrix):
\begin{equation}
\dot{X} = cX + d AX + Xf(AX);
\label{eq:nonlin}
\end{equation}
\noindent where $c,d\in \R$, and $f$ is an arbitrary holomorphic function, with $f(0)=0$. This, for instance, includes spreading--like quadratic terms ($\dot{X} = -cX + d (1-X)A X$).  This equations contains all nonlinear terms in the form $X(AX)^j$, for any $j\in\mathbb{N}_0$. The operator $P=\xi^{-1}\xi$ is an orthogonal projector on the subspace $R$; using $P$, we decompose $X$ in terms of its action on the two subspaces $L$ and $R$ (remember Fig.~\ref{fig:schema}), $X = X_{RR}+X_{RL}+X_{LR}+X_{LL}$: $X_{RR} = PXP$, $X_{RL} = PA(1-P)$ and so on. By definition of $R$ and $L$, $A$ is nonzero only inside $R$: $A=PAP$. Using this, and the McLaurin decomposition of $f$ ($f(z) = \sum_j f_j z^j$) we can prove that $f(AX) = Pf(AX_{RR}) + Pg(AX_{RR})AX_{RL}$. Note that $g$ is another holomorphic function defined as $g(z) = \sum_j f_{j+1} z^j$. We can now decompose Eq.~(\ref{eq:nonlin}) in terms of the four parts of $X$:
\begin{align}
  \dot{X}_{RR} &= cX_{RR} + d  A X_{RR} +  X_{RR} f(AX_{RR});\label{eq:nonlinRR}\\
  \dot{X}_{RL} &= \left\{ c + \left[ d +  X_{RR} g(AX_{RR}) \right] A  \right\} X_{RL};\label{eq:nonlinRL}\\
  \dot{X}_{LR} &= X_{LR} \left[c + f(AX_{RR}) \right];\label{eq:nonlinLR}\\
  \dot{X}_{LL} &= cX_{LL} + X_{LR} g(AX_{RR})A X_{RL}.\label{eq:nonlinLL}
\end{align}
Equation~(\ref{eq:nonlinRR}) is completely restricted to the reduced subspace $R$, and we can use our mapping $\xi$ to send it to $\Rr$, by defining $U = \xi X \xi^{-1}$. The resulting equation is $ \dot{U} = cU + dAU + U f(BU)$: identical to Eq.~(\ref{eq:nonlin}), but living in the reduced space. Once we solve for $U$ (either analytically or numerically, depending on the specific equation), we can go back to $X_{RR}$ by using the inverse transformation.
Once $X_{RR}$ is known, Eq.~(\ref{eq:nonlinRL}), (\ref{eq:nonlinLR}), (\ref{eq:nonlinLL}) and  are just linear, and can be solved with standard techniques~\cite{Dyson1949,Blanes2009,Argeri2014,Tian2014}.
Remarkably, if we assume a simple (and often realistic) initial condition of  $X(0) = \mathbb{I}$, then $X_{LR}, X_{RL}$ are identically zero.
This allows us to write a simple, explicit solution of Eq.~(\ref{eq:nonlin}): $X(t) = \xi^{-1} U(t) \xi + e^{ct} \left(1-\xi^{-1}\xi\right)$.
We have transformed a system of $n^2$ coupled nonlinear differential equations in $n^2$ unknowns, into one of just $r^2$. The gain is dramatic considering that $n$ is the number of nodes (large), while $r$ (the reduced rank) is for most known models very small. Appendix~\ref{sec:nonlin_dyn} contains the detailed solution of Eq.~(\ref{eq:nonlin}).

\section{Symmetries in the space of generative models of networks}
\label{sec:symmetry}

Models in the same class \mymodel{r}{p}, with different metadegrees, may still be the same model in disguise. The action of representations of specific isometry groups in the small ($\Rr$) and large spaces ($\Rn$) may induces isomorphisms between models previously considered as separate physical objects.
Firstly, we consider the isometry group of $\Delta$: $\mbox{Iso}(\Delta)$. The $r\times r$ matrix $Q$ belongs to the representation of $\mbox{Iso}(\Delta)$ in $\Rr$ if it leaves $\Delta$ unchanged: $Q^T \Delta Q = \Delta$. We call them {\itshape internal transformations} because of their action on the metadegrees: $v_{\mu, i}\rightarrow \left. v'\right._{\mu, i} = \sum_\nu Q_{\mu\nu} v_{\nu,i}$. They happen inside a node: they mix its metadegree values, but do not mix metadegree values belonging to different nodes.
Internal transformation leave $A$ unchanged: $A=V\Delta V^T \rightarrow VQ^T \Delta Q V^T = A$. Hence, two models whose metadegrees are connected by an internal transformation are, for all intents and purposes, the same model.
Internal symmetries do modify $B$, though clearly not its spectrum: $B\rightarrow QBQ^{-1}$. We can then choose $Q$ wisely, so that $B$ has the simplest possible form, provide we know the structure of $\mbox{Iso}(\Delta)$. Luckily, we have shown that the known models have either a Euclidean or a Lorentzian signature, whose isometry groups are the most known and studied~\cite{mattila1999geometry}: Respectively, they are the orthogonal group and the $r$-dimensional Lorentz group.
In the Euclidean case ($B=J$) we can go further, as there always exists an orthogonal matrix $Q$ so that $Q B Q^{-1} = Q B Q^T = Q J Q^T$ is diagonal. This means that the rotated metadegrees $VQ^T$ are orthogonal, and since $\Delta = \mathbf{I}$, their norm directly gives the spectrum of $A$.
As a result, whenever the metadegrees are orthogonal (or we can make them so with a rotation in $\Rr$), $i)$ their norms are the nonzero eigenvalues of $A$, $ii)$ the metadegrees are also eigenvectors of $A$. These isometries are of special interest for the understanding of geometrical embeddings of complex networks~\cite{krioukov2010hyperbolic}.
  \begin{figure}\centering
\includegraphics[width=0.45\textwidth]{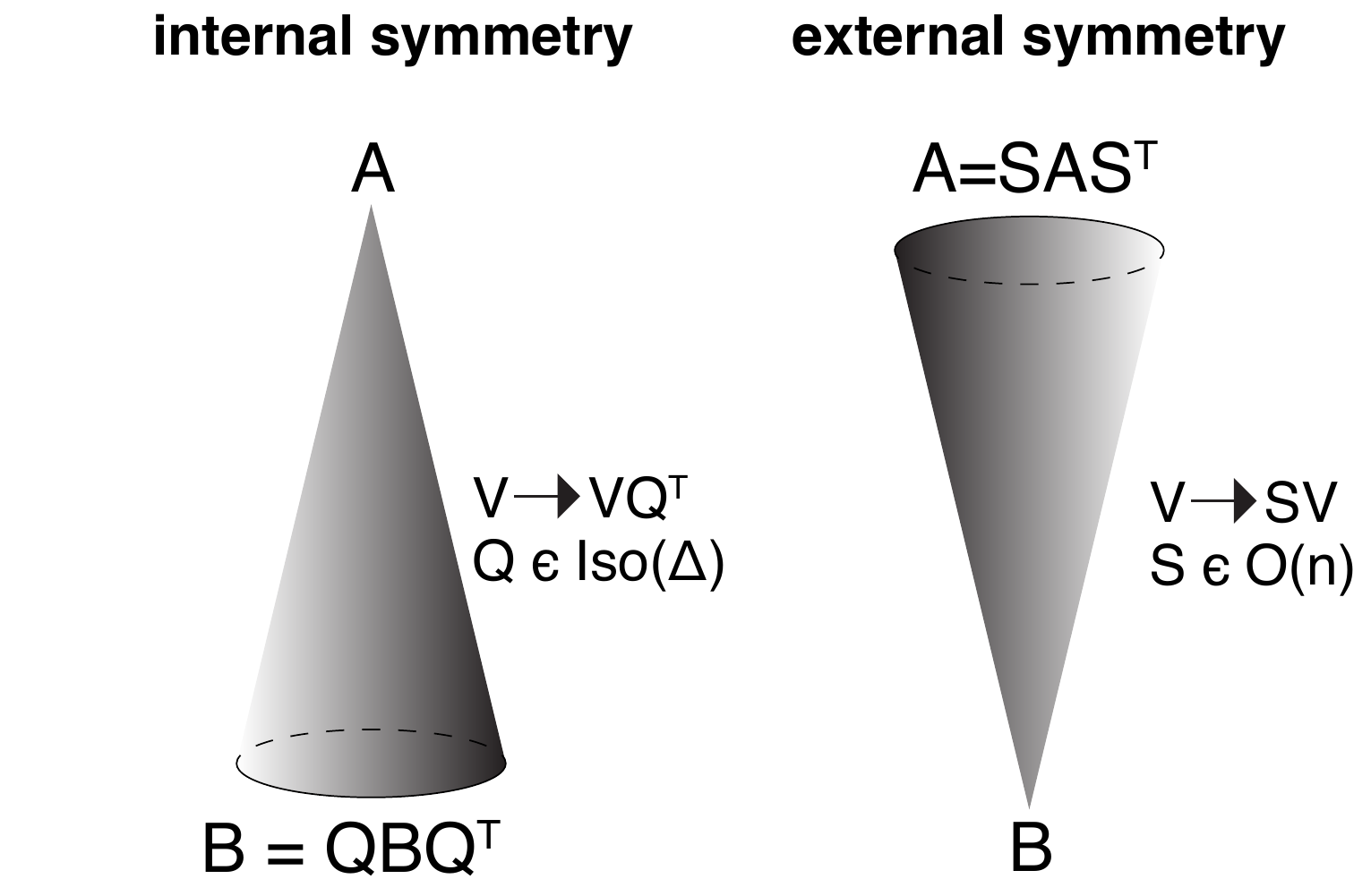}%
 \caption{\label{fig:sym}Schematic representation of internal and external symmetries of the rank-reduction scheme.}
 \end{figure}

We now turn to {\itshape external transformations}. They act on the large space $\Rn$ by mixing the values each metadegree has on the nodes. Opposite to internal transformations, they do mix nodes but do not mix different metadegrees. These transformations comprise the orthogonal group $O(n)$, and act on the metadegrees as follows: $V\rightarrow SV$, for $S\in O(n)$. It is easy to show that they change $A$ through a similarity transformation, while keeping $B$ unchanged.

Summing up, we have found that internal symmetries span different low-rank representations ($B$) of the same model ($A$). External transformations instead span all the models ($A$) that have the same low-rank representation ($B$). Both symmetries, however, preserve the spectrum. Internal and external symmetries are schematically represented in Fig.~\ref{fig:sym}.
We now study the combined action of internal and external symmetry transformations: $V\rightarrow SVQ^T$. In the Euclidean case, this coincides with the singular value decomposition of $V$. In general, it still has far-fetching implications on the nature of models themselves. Within the same class, it allows mapping different models onto each other. Models that are completely different in nature and purposes may have the same properties if they are linked by this symmetry transformation.
As a practical example, we now show that within class \mymodel{2}{1}, the activity-driven model ({\itshape adm} henceforth) can be mapped onto the stochastic blockmodel with two subsets and high inter-subset connectivity ($h>k$) ({\itshape sbm} henceforth). This is quite remarkable if we consider that the former is commonly used to model time-evolving networks with fixed microscopic activity patterns, while the latter applies to static networks featuring mesoscale structures. We start from an {\itshape adm} with fixed activity vector $\Omega$ and number of stubs $m$ (see Eq.~(\ref{eq:adm})). We will land on a {\itshape sbm} featuring two equally-sized subsets, whose degrees $k, h$ will be computed. This means finding the transformation which gives $V_{sbm} = SV_{adm}Q^T$, where $V_{sbm}, V_{adm}$ are the metadegrees of the two models when the metric is in normal form ($\Delta = \mbox{diag}(1,-1)$).
The two-dimensional Lorentz group has one parameter and consists in hyperbolic rotations on the plane; its infinitesimal generator is the first Pauli matrix ($\sigma_1$). Hence, we can span the internal transformations using the hyperbolic angle $\theta$ and by exponentiation of the infinitesimal generator: $Q = e^{\theta\sigma_1}$. The transformation relation, made explicit for each of the two metadegrees, defining $h^{+}=\sqrt{h+k}$ and $h^{-}=\sqrt{h-k}$, becomes 
\begin{equation}
\begin{cases}
(h^{+}+h^{-}) F_1 + (h^{+}-h^{-}) F_2 = e^{-\theta}\sqrt{2 m} SF\\ 
 (h^{+}-h^{-}) F_1 + (h^{+}+h^{-}) F_2 = e^\theta \sqrt{2 m} S\Omega,
\end{cases}
\label{eq:admsbm}
\end{equation}
where the entries of vectors $F_1, F_2$ are $1$ on the first (second) subset, zero otherwise, so that $F_1+F_2=F$. An explicit form of $S$ would then solve Eq.~(\ref{eq:admsbm}). We however only wish to uncover under which conditions such mapping is possible. Thus, we just require that $S$, being orthogonal, preserve standard scalar products in $\Rn$. That fixes the degrees of the {\itshape sbm}: $k=m\pan{a}$, $h=m \sqrt{\pan{a^2}}$. It also fixes the hyperbolic angle of the internal transformation to $\theta = -\frac{1}{4}\log\pan{a^2}$. This demonstrates that we can indeed map the {\itshape adm} onto a {\itshape sbm}, and that the mapping fixes the parameters of such {\itshape sbm}, and also fixes the gauge induced by internal transformations. 

As a last comment, we point out that our classification scheme could be used to generate new models. Traditionally, models have been designed ad hoc to investigate specific network properties.
Now, for each class \mymodel{r}{p}, one could enumerate all the potential models it contains, by means of a systematic characterization of its internal and external symmetries. Some of these models could feature previously unrepresented and unstudied network properties.

\section{Product of generative models of networks}
\label{sec:promodels}

The stochastic blockmodel, which we have already examined, features a rank that equals the number of subsets (or communities). If one needs a stochastic blockmodel with many subsets (fine partition), the rank may then be large. This seems to go against  our claim that our rank-reduction is powerful because the typical required rank is small, and happens because of the presence of multiscale structures. We posit that if a model features organizational scales above node-node correlations (clustering, communities, multipartiteness and so on) its rank needs to be large. It might even need to scale with the size of the system. We introduce here a new operation that allows us to overcome this apparent drawback by reducing high-rank structures to low rank components. This will allow to extend our classification to models that exhibit complex mesoscale features and to apply the full power of our machinery to them.

We define the tensor product of two models as the tensor (Kronecker) product of their matrix representation. We start from a model $A_1 = V_1 \Delta_1 V_1^T$, with $n_1$ nodes and rank and signature $r_1, p_1$, and a model $A_2 = V_2 \Delta_2 V_2^T$ with $n_2, r_2, p_2$ respectively. The product model $A = V\Delta V^T$ is
\begin{eqnarray}
\nonumber
A = A_1 \otimes A_2 = (V_1 \Delta_1 V_1^T)\otimes (V_2 \Delta_2 V_2^T) = \\
(V_1 \otimes V_2) (\Delta_1 \otimes \Delta_2) (V_1 \otimes V_2)^T.
\label{eq:tensor}
\end{eqnarray}
Equation~(\ref{eq:tensor}) shows that the metadegree matrix and the metric are simply the tensor product of the original matrices: $V = V_1 \otimes V_1^T$ and $\Delta = \Delta_1 \otimes \Delta_2$. The resulting network has $n = n_1 n_2$ nodes, rank $r = r_1 r_2$, and signature
\begin{equation}
p = r \left( 1 - \frac{p_1}{r_1} - \frac{p_2}{r_2} + 2  \frac{p_1}{r_1}  \frac{p_2}{r_2}  \right).
\label{eq:sign_product2}
\end{equation}
Appendix~\ref{sec:prosign} contains the proof of Eq.~(\ref{eq:sign_product2}).

A model which is a tensor product of two smaller models is thus completely defined and characterized by its factors. This is a further simplification as high-rank models can actually be studied in terms of their low-rank factors. The properties of the spectrum, linear and nonlinear dynamics completely follow from the study of the factors. For instance, the spectrum of $A$ is composed of all the possible products of one eigenvalue of $A_1$ with one eigenvalue of $A_2$.

In general, the product decomposition of models simplifies the classification and treatment of high-rank models. Any model whose rank $r$ is not a prime number can be decomposed into the product of $\tau$ lower-rank models, being $\tau$ the number of factors of $r$:
\begin{equation}
\mymodelm{r}{p} = (\mymodelm{r_1}{p_1})\otimes (\mymodelm{r_2}{p_2} )\otimes \cdots \otimes (\mymodelm{r_\tau}{p_\tau} ),
\label{eq:product_decomposition}
\end{equation}
provided
\begin{align}
r &= \prod_{j=1}^\tau r_j, \label{eq:product_rank} \\
p &= \frac{1}{2} \left[ r + \prod_{j=1}^\tau (2p_i - r_i)  \right]. \label{eq:produc_sign}\\\nonumber
\end{align}

Appendix~\ref{sec:prosign} contains the proof of Eq.~(\ref{eq:produc_sign}).

\subsection*{Model entanglement}

While a decomposition of the form in Eq.~(\ref{eq:product_decomposition}) always exists, Eq.~(\ref{eq:product_rank},\ref{eq:produc_sign}) are not sufficient for Eq.~(\ref{eq:product_decomposition}) to hold. In other words, a model may or may not be decomposed by a specific tensor product respecting Eq.~(\ref{eq:product_rank},\ref{eq:produc_sign}).

As an example, let us consider a generic model in \mymodel{4}{2}. We can decompose it in two ways: as the product of a degree-degree correlated configuration model and and activity-driven model $\mymodelm{2}{2}\otimes \mymodelm{2}{1}$, or as the product of two activity-driven models: $\mymodelm{2}{1} \otimes \mymodelm{2}{1}$. Let us choose the former. The general form of a model in \mymodel{4}{2} is $A = v_1 v_1^T + v_2 v_2^T - v_3 v_3^T - v_4 v_4^T$. As we proved before, the four metadegrees span a 4-dimensional space isomorphic to $\R^4$. We change the basis of such space to highlight a $\R^2\otimes \R^2$ structure: $v_1 = e_1 \otimes f_1$, $v_2 = e_1 \otimes f_2$, $v_3 = e_2 \otimes f_1$, $v_4 = e_2 \otimes f_2$. By substituting it into the expression of the model we get
\begin{eqnarray}
\nonumber
A = \sum_{a,b,c,d=1}^2 \eta_{ab} \delta_{cd} (e_a \otimes f_c) (e_b \otimes f_d)^T = \\
\sum_{a,b,c,d=1}^2 \left( \eta_{ab} e_a e_b^T \right) \otimes \left( \delta_{cd} f_c f_d^T \right),
\label{eq:rank4dec}
\end{eqnarray}
where $\eta = \mbox{diag}(1,-1)$ is the 2-dimensional Lorentz metric. Equation~(\ref{eq:rank4dec}) highlights the decomposition into a model \mymodel{2}{2} ($\delta_{cd} f_c f_d^T$) and a model \mymodel{2}{1} ($\eta_{ab} e_a e_b^T$). However, if we choose a change of basis that mixes the metadegrees in some other way the product might not factorize. This is equivalent to the entangled states in quantum mechanics. A model is separable with respect to two potential factors if it can be expressed in terms of a product of the two factors. It is entangled if this is not true. Inspired by the quantum mechanical analogy we state that a model $A$ is separable with respect to the product $A_1 \otimes A_2$ if and only if
\begin{equation}
A = \frac{\tr_2 A \otimes \tr_1 A}{\tr A},
\end{equation}
where $\tr_1$ ($\tr_2$) is the trace of $A$ in the space where $A_1$ ($A_2$) lives.

\subsection*{Creation of new multiscale structures}

We have already shown that the stochastic blockmodel with $c$ subsets, $n_c$ nodes per subset ($n_c = n/c$), belongs to either class \mymodel{c}{c} or \mymodel{c}{1}. The tensor decomposition provides a natural way to simplify it further, and then generalize it:
\begin{equation}
\mbox{sbm}(c=c, n_c=n_c) \equiv \mbox{sbm}(c=c, n_c=1)\otimes \mbox{cm}
\label{eq:sbm_tensor}
\end{equation}
\noindent being cm the configuration model.
The stochastic blockmodel is then the product of the same model but with only one node per subset, and a configuration model. The former describes between-subset connectivity, and the latter within-subset connectivity. From this relation, a very important interpretation of the product becomes apparent: The tensor product of models is deeply connected with a multiscale structure of the network. Once we explicit this through Eq.~(\ref{eq:sbm_tensor}), we immediately understand that we do not need to stick to the configuration model for the within-subset connectivity: we can plug in whatever model we want. For instance, we can define a stochastic blockmodel with an activity-driven model within the subsets simply like this:
\begin{equation}
\mbox{sbm}(c=c, n_c=n_c) = \mbox{sbm}(c=c, n_c=1)\otimes \mbox{adm}.
\label{eq:sbm_tensor_adm}
\end{equation}
This model will have rank $r=2c$ and signature $p=c$ regardless of the signature of the stochastic blockmodel.

\subsection*{Kronecker graphs}

The product of models defined above bears a clear resemblance to Kronecker graphs~\cite{Leskovec2005,Leskovec2007,Leskovec2010,Kim2010,Seshadhri2013}. Including Kronecker graphs into our classification scheme is important because of the nontrivial properties they exhibit, like clustering and $k$-core organization~\cite{Robles2016}.
As we have seen, mesoscale structures in the network likely require a high rank, possibly scaling as the size of the system. This would seem as an apparent drawback of our theory. Including Kronecker graphs using model products solves that. Even if the rank of the full model scales with the number of nodes, the rank of the base model remains small, and this is the only thing we need to solve it.
We set as base of the Kronecker graph a small-rank model $\hat{A}$ with $\hat{n}, \hat{r}, \hat{p}$. For a definition of base (or initiator) of a Kronecker graph see Ref.~\cite{Leskovec2005,Mahdian2007}. The Kronecker model $A$ is then the tensor power
\begin{equation}
A = \hat{A}^{[\tau]} = \underbrace{\hat{A} \otimes \hat{A} \otimes \cdots \otimes \hat{A}}_{\tau\, times},
\end{equation}
with $\tau$ being the power of the model: $\tau = \log n / \log \hat{n}$. The metadegrees and metric of the full model are $V = \hat{V}^{[\tau]}, \Delta = \hat{\Delta}^{[\tau]}$. The rank and signature of the full model are
\begin{align}
r &= \hat{r}^\tau = n^{\log\hat{r}/\log\hat{n}} \label{eq:kroandamento}\\
p &= \frac{ \hat{r}^\tau + (2 \hat{p}-\hat{r})^{\tau } } {2} \label{eq:kroandamentop} 
\end{align} 
Equation~(\ref{eq:kroandamento}) shows that the rank increases with the size of the system. Notwithstanding, the model is easy to solve because one needs only solve its base $\hat{A}$. Appendix~\ref{sec:prosign} contains the proof of Eq.~(\ref{eq:kroandamentop}).

\section{An application: communicability}

Finally, to highlight the practical advance we can directly drawn from our proposal, let us comment on a simple application of this reduction to compute a magnitude that directly depends on the matrix form of a network, the communicability.
Communicability $C_{ij}$ accounts for all the possible walks, which are generalized paths allowing for repetition of nodes and edges --~of any length (weighted in decreasing order of length)~-- that join $i, j$, and measures how easily information flows between the two nodes~\cite{Estrada2008,Estrada2012}.
In its most common form, walks have a penalization equal to the inverse of the factorial of their length, so that communicability has the following closed form: $C=e^A$.
As a practical application of our theory, we now study the communicability of models. In particular, we examine the relationship between local connectivity ($A$) and nonlocal connectivity ($C$).

Inserting Eq.~(\ref{eq:decomposition}) in the definition of the communicability, we get
\begin{align}
C &= \sum_{m=0}^\infty \frac{1}{m!} \left(V\Delta V^T\right)^m = 1 + V\Delta \sum_{m=1}^\infty \frac{1}{m!}B^{m-1} V^T \nonumber \\
C &= 1 + V\Delta B^{-1} \left( e^B -1  \right) V^T = 1 + V\Psi V^T. \label{eq:communicability}
\end{align}
where $\Psi = \Delta B^{-1} \left(e^B-1\right)$, and $\Psi\in\R^{r,r}$. $C$ has maximal rank ($n$), as it is the exponential of $A$. However, rank-reduction returns a very simple form. The high rank structure is simply represented by the identity matrix. Intuitively, this is the exponentiation of the large trivial subspace with eigenvector zero.
The nontrivial part of the exponentiation ($\Psi$) occurs within the small space, as Eq.~(\ref{eq:communicability}) shows. This means that communicability can be rank-reduced, and computed in the small space.
Notably, $\Psi$ is also a well-behaved metric in $\Rr$ (it is symmetric), and $V\Psi V^T$ is itself a network model (compare it with Eq.~(\ref{eq:decomposition})), with rank equal to $r$, as $\Psi$ is nonsingular.
This implies that the communicability $C$ of a model $A$ in \mymodel{r}{p} is itself a model with the same rank $r$. What about the signature of this new model? In principle, communicability may fall in any class \mymodel{r}{p'}. If $B$ and $\Delta$ commute, then it is easy to prove that $p'=p$. In any other case, communicability may change universality class with respect to the original model.
Commutation between $B, \Delta$ is guaranteed when the original model is Euclidean ($p=r$), or, in general, whenever the metadegrees corresponding to different signs in $\Delta$ are orthogonal. In other words, $v_\mu \cdot v_\nu \not=0$ only if $\Delta_{\mu\mu} = \Delta_{\nu\nu}$ (with $\Delta$ in normal form).

Summing up, the practical implications are the following. i) You can express the nonlocal connectivity patterns (communicability) of any network model, in terms of the local connectivity (adjacency matrix) of a model with the same rank as the original one. ii) In general, the universality class of the communicability and the one of the original model are different ($p'\not=p$). This means that nonlocal connectivity patterns exhibit a topology that is qualitatively different from local ones: communicability is a nontrivial emergent property of the local model. iii) Rank-reduction allows you to build network ensembles with given communicability. We do not explore this in the present paper, but we think it can lead to promising practical implications. iv) In some specific cases (Euclidean models, models that do not mix different signature subspaces) the communicability is in the same universality class of the original model, hinting at the fact that these models might exhibit some forms of scale invariance.

To conclude this application, we explicitly compute the communicability of the configuration model and the activity-driven model.
\begin{equation}
C^{(cm)}_{ij} = \delta_{ij} + \frac{\pan{k}}{\pan{k^2}} \left( e^{   \pan{k^2}/\pan{k} } - 1  \right) \frac{k_i k_j}{n \pan{k}};
\end{equation}
the communicability of the configuration model is proportional to its adjacency matrix. We now turn to the activity-driven model (\mymodel{2}{1}), with metadegrees as in Eq.~(\ref{eq:adm}), and $\Delta = \mbox{diag}(1,-1)$. Given that $v_1 \cdot v_2 = 0$, the communicability of this model is also \mymodel{2}{1}. We first compute $\Psi$:
\begin{equation}
\Psi_{adm} = \frac{1}{m}
\begin{pmatrix}
\frac{ e^{m (\sqrt{\pan{a^2}} +\pan{a} ) }-1 }{ \sqrt{\pan{a^2}} +\pan{a} } & 0 \\
0 & \frac{ e^{-m (\sqrt{\pan{a^2}} - \pan{a} ) }-1 }{ \sqrt{\pan{a^2}} -\pan{a} } 
\end{pmatrix}.
\end{equation}
For brevity we define 
\begin{align}
\psi_1 &= \frac{ e^{m (\sqrt{\pan{a^2}} +\pan{a} ) }-1 }{ \sqrt{\pan{a^2}} +\pan{a} } \\
\psi_2 &= \frac{ e^{-m (\sqrt{\pan{a^2}} - \pan{a} ) }-1 }{ \sqrt{\pan{a^2}} -\pan{a} } .
\end{align}
We now can write the communicability matrix for the activity-driven model:
\begin{align}
C^{(adm)}_{ij} &= \delta_{ij} + \frac{1}{2n} \left[ (\psi_1+\psi_2) \left( \sqrt{\pan{a^2}} + \frac{a_i a_j}{\sqrt{\pan{a^2}}} \right) + \right.\\
 &\left.+ (\psi_1-\psi_2) \left( a_i + a_j \right)   \right].
\end{align}
Notably, the communicability of the basic activity-driven model exhibits the diagonal terms $FF^T$ and $\Omega\Omega^T$ like the adjacency matrix of the simplicial activity-driven model (see Appendix~\ref{seq:petri}).

\section{Conclusion}

We have proposed a linear algebraic methodology for classifying generative network models. By using the--rank reduction of the matrix representation of generative models of networks, we have derived many of their properties, and the properties of the dynamical processes on top of them.
We did this by solving the generic model, so that our results apply algorithmically to all specific cases, with a dramatic decrease in the complexity of the analytic and numerical calculations involved. Our scheme includes models featuring many properties that are needed to study real datasets. This are both local properties, like node degree or activity, as well as more complex nonlocal features, like mesoscale structures.
Finally, we have shown how the geometrical properties of our scheme can be used to devise new models, extend and make connections between existing ones.

Some network structures still elude our rank-reduction scheme. We do not account for regular topologies (like lattices). Neither we cover finite-size effects ($n$ far from the thermodynamic limit). Time-evolving network models featuring temporal correlations are also not included at the moment, but we are confident that future formulations will extend to them. For example, temporal correlations could emerge out of tensor representations of models~\cite{DeDomenico2013,gauvin2014detecting}, and finite-size effects as perturbations of the spectrum of $A$ with diagonal matrices~\cite{Arbenz1988b}. These are just ideas we are considering for a future work.


\section*{Acknowledgments}
We thank Mason Porter and Michele Re Fiorentin for the useful discussions. AA acknowledges support by Ministerio de Econom\'{\i}a y Competitividad (grant FIS2015-71582-C2-1, PGC2018-094754-B-C21), Generalitat de Catalunya (grant 2017SGR-896), and Universitat Rovira i Virgili (grant 2017PFR-URV-B2-41), ICREA Academia and the James S.\ McDonnell Foundation (grant \#220020325).

\newpage

\section*{Appendices}

\appendix

\section{The activity-driven model}
\label{sec:adm}

We start from the generalized activity-driven model presented in~\cite{Alessandretti2017}, of which the original formulation of~\cite{Perra2012} is a special case.

The model fixes two features for each node: the activity rate $a_i$ and the actractiveness $b_i$. A node activates at a rate $a_i$ and, when active, it links to $m$ other nodes chosen with probability proportional to their actractiveness $b_j$. The expected value of each link thus is
\begin{equation}
	A_{ij} = \frac{m}{n \pan{b}} \left( a_i b_j + a_j b_i  \right).
	\label{eq:adm_mat_gen_ij}
\end{equation}
We arrange activity rates in vector $\Omega$: $\Omega_i = a_i$. We arrange attractiveness values in vector $\Sigma$: $\Sigma_i = b_i / \pan{b}$. This results in
\begin{equation}
A = \frac{m}{n} \left( \Omega \Sigma^T + \Sigma \Omega^T \right),
\label{eq:adm_mat_gen}
\end{equation}
proving that the activity-driven model has rank $r=2$. We now diagonalize the metrics and get to the metadegrees. We define
\begin{eqnarray}
	U_1 &= \left( \frac{m}{2 n \pan{b}} \right)^{\frac{1}{2}} \left( \frac{\pan{b^2}}{\pan{a^2}} \right)^{\frac{1}{4}} \left[  \Omega + \left( \frac{\pan{b^2}}{\pan{a^2}} \right)^{-\frac{1}{2}} \Sigma  \right]; \\
	U_2 &= \left( \frac{m}{2 n \pan{b}} \right)^{\frac{1}{2}} \left( \frac{\pan{b^2}}{\pan{a^2}} \right)^{\frac{1}{4}} \left[  \Omega - \left( \frac{\pan{b^2}}{\pan{a^2}} \right)^{-\frac{1}{2}} \Sigma  \right];
\end{eqnarray}
These vector diagonalize the expression in Eq.~(\ref{eq:adm_mat_gen}):
\begin{equation}
A = \frac{m}{n} \left( U_1 U_1^T - U_2 U_2^T \right).
\end{equation}
$U_1, U_2$ are metadegrees of the model and they show the signature $p=1$. In addition, their specific normalization diagonalizes $B$, as $U_1 \dot U_2 = 0$. In the formalism of Sec.~\ref{sec:symmetry} we have used the degeneracy induced by internal transformation to choose metadegrees which automatically diagonalize $B$. Because of this, the two nonzero eigenvalues of $A$ are $(m/n) \left\Vert U_1 \right\Vert^2$ and $-(m/n) \left\Vert U_2 \right\Vert^2$, i.e., 
\begin{equation}
\frac{m}{\pan{b}} \left(\pan{ab} \pm \sqrt{\pan{a^2}\pan{b^2}} \right).
\end{equation}

Setting $b_i=1$ recovers original version of the activity-driven model:
\begin{equation}
A_{ij} = \frac{m}{n} \left( a_i + a_j  \right).
\label{eq:adm_mat_ij}
\end{equation}
\begin{equation}
m \left( \pan{a} \pm \sqrt{\pan{a^2}} \right).
\end{equation}
Compare the above result with Eq.~(4) of Ref.~\cite{Perra2012}.

\subsection{Simplicial activity-driven model}
\label{seq:petri}

In the version of the activity-driven model introduced in Ref.~\cite{Petri2018} an active node builds a clique (complete subgraph) with other $q-1$ random nodes. It thus generates a $q$-clique. $q$ is sampled from a given distribution. The value $q=2$ recovers the standard activity-driven model (with $m=1$).

One can show that the simplicial structure induces a correction to Eq.~(\ref{eq:adm_mat_ij}):

\begin{equation}
A_{ij} =  \frac{\pan{q}-1}{n} \left(a_i + a_j\right) + \frac{\pan{a(q-1)(q-2)}}{n}(1-a_i)(1-a_j).
\end{equation}


The matrix form of $A$ is
\begin{eqnarray}
\nonumber
A &= \frac{\pan{q}- \pan{ a (q-1)(q-2)}-1}{n} \left( \Omega F^T + F \Omega^T \right) +\\
& \frac{\pan{a (q-1)(q-2)}}{n} \left( FF^T + \Omega\Omega^T \right),
\end{eqnarray}
where we remind that $F$ is the $n$-dimensional vector of ones: $F_i = 1$. From this expression the (nonnormal) metric is 
\begin{equation}
\Delta = \begin{pmatrix}
\frac{\pan{a (q-1)(q-2)}}{n} & \frac{\pan{q}- \pan{ a (q-1)(q-2)}-1}{n} \\
\frac{\pan{q}- \pan{ a (q-1)(q-2)}-1}{n} & \frac{\pan{a (q-1)(q-2)}}{n} \\
\end{pmatrix}.
\end{equation}
The eigenvalues of this matrix are
\begin{align}
	\mbox{eigenvalue}_1 &= \frac{\pan{q}-1}{n}; \\
	\mbox{eigenvalue}_2 &= \frac{1-\pan{q} +2 \pan{a(q-1)(q-2)} }{n}.
\end{align}
The first eigenvalue is always positive, given that $q\geq 2$. The second eigenvalue is positive iff $\pan{a (q-1)(q-2)} > (\pan{q}-1)/2$. When this relation among $\pan{aq^2}, \pan{aq}, \pan{a}, \pan{q}$ holds, the model is \mymodel{2}{2}. Otherwise, it is \mymodel{2}{1}. When instead $\pan{a (q-1)(q-2)} = (\pan{q}-1)/2$, the model is a rank-1 configuration model.
This proves that the model is characterizable and solvable for a generic joint distribution of $q, a$. For simplicity, we now carry out the calculation in the simple case of constant $q$. The form of $A$ is
\begin{eqnarray}
\nonumber
A &= \frac{q-1}{n} \left( 1-\frac{\pan{a}}{2a_c} \right) \left(\Omega F^T + F \Omega^T  \right) +\\
&  \frac{q-1}{n} \frac{\pan{a}}{2a_c} \left( FF^T + \Omega\Omega^T \right),
\label{eq:asimpl_sim}
\end{eqnarray}
where we explicitly highlighted a critical activity value: $a_c = \left[2(q-2)\right]^{-1}$. Equation~\ref{eq:asimpl_sim} clearly shows that $a_c$ discriminates between a regime where the Lorentzian part dominates and another where the Euclidean part dominates. Indeed for high average activity ) $\pan{a} > a_c$ the model is Euclidean (\mymodel{2}{2}). For low average activity $\pan{a} < a_c$ the model is Lorentzian (\mymodel{2}{1}) like the standard activity-driven model. Instead, when $\pan{a} = a_c$, the model collapses onto a rank $r=1$ configuration model with node degree proportional to $1+a_i$.

Now that the classification is complete, one can compute the eigenvalues of $A$ by computing $B$, getting the same threshold condition as in Ref.~\cite{Petri2018}.
%
%

\section{The configuration model with degree-degree correlations}
\label{sec:ddcorr}

The rank $r=2$ euclidean class \mymodel{2}{2} is the simplest (min rank, Euclidean) model featuring degree--degree correlations. It exhibits arbitrary assortative or disassortative behavior.
The general (diagonal) form of this model is
\begin{equation}
A = \frac{1}{n} \left( uu^T + vv^T\right),
\end{equation}
where $u, v$ are its two metadegrees. We have made the dependence of metadegrees on system size explicit. We change the metadegree basis so that one of them is the degree vector $k = \pan{u} u + \pan{v} v$ is a metadegrees ($k_i$ is the degree of node $i$). To do this we set $v = \left( k=\pan{u} u\right) / \sqrt{\pan{k}-\pan{u}^2}$, provided the expression inside the root is positive. The matrix representation is now
\begin{equation}
A = \frac{1}{n} \frac{1}{\pan{k}-\pan{u}^2} \left[ \pan{k}uu^T + kk^T - \pan{u} \left( uk^T + ku^T \right)  \right].
\label{eq:assoA}
\end{equation}
Following Ref.~\cite{dynpas2001} we measure the assortativeness as 
\begin{equation}
\mca{A}(k) = \pan{\frac{d k^{(nn)}}{d k}},
\end{equation}
where $k^{(nn)}_i = \sum_{j!\not= i} A_{ij}k_j / k_i$ is the expected degree of the neighbors of $i$. Using Eq.~(\ref{eq:assoA}), we compute
\begin{equation}
k^{(nn)} = ak + bu,
\end{equation}
with 
\begin{align}
	a &= \frac{\pan{k^2}-\pan{u}\pan{uk}}{\pan{k}-\pan{u}^2}; \\
	b &= \frac{-\pan{u}\pan{k^2}+\pan{k}\pan{uk}}{\pan{k}-\pan{u}^2}. \\
\end{align}
This results in
\begin{equation}
\mca{A}(k) = \frac{b}{k} \left( \frac{du}{dk} - \frac{u}{k}  \right).
\end{equation}
Assuming a constant assortativity $\mca{A}(k) = \mca{A}$, one can solve the differential equation in $u(k)$, getting
\begin{equation}
u(k) = \left( u_0 + \frac{\mca{A}}{b} \log k  \right) k,
\label{eq:ukasso}
\end{equation}
where $u_0 = u(1)$. From this we see that the model \mymodel{2}{2} allows any arbitrary value of assortative ($\mca{A}>0$) or disassortative ($\mca{A}<0$) behavior, by tuning the dependence of the second metadegree $u$ on the degree. Furthermore, Eq.~(\ref{eq:ukasso}) tells us that for no assortativity ($\mca{A}=0$), the model reduces to \mymodel{1}{1} --~the rank $r=1$ configuration model~--, because $u = u_0 k$ means that $u, k$ are no longer linearly independent. Instead, we see that degree--degree correlations arise from a logarithmic correction to the $\mca{A}=0$ solution.
The free parameter $u_0$ is fixed by consistency, as $b$ contains moments of $u,k$ whose value must be compatible with the solution. 

\section{The stochastic blockmodel}
\label{sec:sbm}

The matrix representation of the model is
\begin{align}
A &= \frac{ck}{n} \sum_{\mu=1}^c F_\mu F_\mu^T + \frac{ch}{n (c-1)} \sum_{\mu\not= \nu} F_\mu F_\nu^T; \\
A &= \frac{c}{n} \left\{ \left[ k   - \frac{h}{ (c-1)}   \right] \sum_{\mu=1}^c F_\mu F_\mu^T + \frac{h}{(c-1)} \sum_{\mu, \nu=1}^c F_\mu F_\nu^T \right\}; \\
A &= \frac{c}{n} \left\{ \left[ k   - \frac{h}{ (c-1)}   \right] \sum_{\mu=1}^c F_\mu F_\mu^T + \frac{h}{(c-1)} F F^T \right\}; 
\end{align}
where $F_\mu$ has entries equal to one corresponding to subset $\mu=1,\cdots, c$, zero otherwise. These vectors $\{\frac{c}{n} F_1, \frac{c}{n} F_2, \cdots \frac{c}{n} F_c  \}$ are linearly independent. This proves the rank $r=c$. We choose them as metadegrees. The metric (not in normal form) is
\begin{align}
	\Delta_{\mu\mu} &= k; \\
	\Delta_{\mu\nu} &=  \frac{h}{c-1}, \mbox{for } \mu\not= \nu.
\end{align}
This matrix has one eigenvalue $k+h$ with multiplicity $c-1$, which is always positive. It also has one eigenvalue $k-h/(c-1)$, with multiplicity one. Hence, if $k>h/(c-1)$, the signature is $p=r=c$ (Euclidean). If $k<h/(c-1)$, the signature is $p=1$ (Lorentzian).

\section{Nonlinear dynamics}
\label{sec:nonlin_dyn}

We detail the solution of Eq.~(\ref{eq:nonlin}), in terms of the solution of the system in Eq.~(\ref{eq:nonlinRR},\ref{eq:nonlinRL},\ref{eq:nonlinLR},\ref{eq:nonlinLL}).
Equation~(\ref{eq:nonlinRR}) is the only one containing the nonlinearity, but it is fully contained in the small space $\Rr$. We solve that for $X_{RR}$ (either analytically or numerically) as explained in the main paper ($X_{RR}(t) = \xi^{-1} U(t) \xi$); the other equations then become linear. Therefore they are always solvable~\cite{Blanes2009,Tian2014}, using series expansions like Dyson's~\cite{Dyson1949}.

Before proceeding any further, we recall the definition of Dyson's time-ordering operator $\mca{T}$:
\begin{eqnarray}
\nonumber
\mca{T} \left[ X(t_1) X(t_2) \right] = \theta(t_1-t_2) X(t_1)X(t_2) + \\
\theta(t_2-t_1) X(t_2)X(t_1), 
\end{eqnarray}
with $\theta$ being Heaviside's step function. We also define Dyson's time-ordered exponentiation:
\begin{align}
&\mca{T} \exp \left( \int_0^t \de{s} X(s) \right)= \\
&\sum_{m=0}^\infty \frac{1}{m!} \int_{0}^{t}\de{s_1} \de{s_2}\cdots \de{s_m} \mca{T} \left[ X(s_1)X(s_2) \cdots X(s_m) \right] =  \nonumber \\
& \sum_{m=0}^\infty  \int_{0}^{t}\de{s_1} \int_{0}^{s_1} \de{s_2}\cdots \int_{0}^{s_{m-1}} \de{s_m}  X(s_m)\cdots X(s_2) X(s_1).
\end{align}

We now start from Eq.~(\ref{eq:nonlinRL}). It is a nonautonomous linear system with unknown $X_{RL}$. We remind that $X_{RR}$ is no longer unknown, as it is fully determined by Eq.~(\ref{eq:nonlinRR}). Its solution in terms of Dyson's series is
\begin{align}
\nonumber
&X_{RL}(t) =  \\
&\mca{T} \exp \left\{  \int_0^t \de{s}  \left( c + dA + X_RR(s) g[AX_{RR}(s)] A \right)   \right\} X_{RL} (0).
\label{eq:nonlinsolved_RL}
\end{align}
The same applies for 
\begin{equation}
X_{LR}(t) =  X_{LR} (0) \mca{T} \exp \left\{  \int_0^t \de{s}  \left( d + f[AX_{RR}(s)]  \right) \right\} .
\label{eq:nonlinsolved_LR}
\end{equation}
Finally, once we have $X_{RL}$ and $X_{LR}$, Eq.~(\ref{eq:nonlinLL}) gives $X_{LL}$ as a simple autonomous nonhomogeneous linear system:
 \begin{eqnarray}
 \nonumber
& X_{LL}(t) = e^{ct} X_{LL}(0) +  \\
& e^{ct}\int_0^t \de{s} e^{-cs} X_{LR}(s) g[AX_{RR}(s)] A X_{RL}(s).
 \label{eq:nonlinsolved_LL}
 \end{eqnarray}
 
 Enforcing as initial condition something proportional to the identity operator simplifies the calculation. $X(0) = \epsilon\in \R$, means $X_{RR}(0) = X_{LL}(0) = \epsilon$, and $X_{RL}(0) = X_{LR}(0) = 0$.
 Both Eq.~(\ref{eq:nonlinsolved_RL}) and Eq.~(\ref{eq:nonlinsolved_LR}) are now identically zero: $X_{RL}(t) = X_{LR}(t) = 0$, $\forall t$. Equation~(\ref{eq:nonlinsolved_LL}) reduces to 
 \begin{equation}
 X_{LL}(t) = \epsilon e^{ct}.
 \end{equation}
 
 Finally, the full solution is $X(t) = X_{RR}(t)+ X_{LL}(t)$:
 \begin{equation}
 X(t) =  \xi^{-1}U(t)\xi + \epsilon e^{ct} \left( 1-\xi^{-1}\xi \right).
 \end{equation}
 
 \section{Signatures of product models}
 \label{sec:prosign}
 
 We prove Eq.~(\ref{eq:sign_product2}). The eigenvalues of the product model $A=A_1\otimes A_2$ are all the possible products of one eigenvalue of the first factor model ($A_1$), with one eigenvalue of the second factor ($A_2$). $A_1$ has rank $r_1$ and $p_1$ positive eigenvalues. $A_2$ has rank $r_2$ and $p_2$ positive eigenvalues. A positive eigenvalue of $A$ must be either the product of a positive eigenvalue of $A_1$ and a positive eigenvalue of $A_2$, or the product of a negative eigenvalue of $A_1$ and a negative eigenvalue of $A_2$. Hence, $p = p_1 p_2 + (r_1-p_1) (r_2 - p_2)$. By collecting $r=r_1 r_2$, one gets to Eq.~(\ref{eq:sign_product2}).
  
 We prove Eq.~(\ref{eq:produc_sign}) for model in Eq.~(\ref{eq:product_decomposition}). The negative eigenvalues of the product model are the product of an odd number of negative eigenvalues coming from the factor models. We define $q_j = r_j - p_j$ as the number of negative eigenvalues of the $j$-th factor, and $q=r-p$ as the number of negative eigenvalues of the product model. We can make explicit the multiplicity of eigenvalues by writing
 \begin{equation}
 r = p+q = \prod_{j=1}^{\tau} r_j = \prod_{j=1}^{\tau} \left( p_j + q_j \right).
  \label{eq:krf1}
 \end{equation}
 The expansion of the product would show all the possible ways to pick positive and negative eigenvalues from the factors, and the respective multiplicities.
 We consider also
 \begin{equation}
 \prod_{j=1}^{\tau} \left( p_j - q_j \right).
 \end{equation}
 In this case the factors of the expanded product are positive if the corresponding multiplicity refers to positive eigenvalues, and they are negative if the corresponding multiplicity refers to negative eigenvalues. Hence,
  \begin{equation}
 \prod_{j=1}^{\tau} \left( p_j - q_j \right) = p-q.
 \label{eq:krf2}
 \end{equation}
 Combining Eq.~(\ref{eq:krf1}) and Eq.~(\ref{eq:krf2}) we compute $p$ and get to Eq.~(\ref{eq:produc_sign}).

 Finally, Eq.~(\ref{eq:kroandamentop}) is simply the special case of Eq.~(\ref{eq:produc_sign}) when all the factors have same rank and signature.



\begin{thebibliography}{10}
	
	\bibitem{borge2016dynamics}
	Javier Borge-Holthoefer, Nicola Perra, Bruno Gon{\c{c}}alves, Sandra
	Gonz{\'a}lez-Bail{\'o}n, Alex Arenas, Yamir Moreno, and Alessandro
	Vespignani.
	\newblock The dynamics of information-driven coordination phenomena: A transfer
	entropy analysis.
	\newblock {\em Science advances}, 2(4):e1501158, 2016.
	
	\bibitem{halu2019}
	Arda Halu, Manlio De~Domenico, Alex Arenas, and Amitabh Sharma.
	\newblock The multiplex network of human diseases.
	\newblock {\em npj Systems Biology and Applications}, 5(1):15, 2019.
	
	\bibitem{stelzl2005human}
	Ulrich Stelzl, Uwe Worm, Maciej Lalowski, Christian Haenig, Felix~H Brembeck,
	Heike Goehler, Martin Stroedicke, Martina Zenkner, Anke Schoenherr, Susanne
	Koeppen, et~al.
	\newblock A human protein-protein interaction network: a resource for
	annotating the proteome.
	\newblock {\em Cell}, 122(6):957--968, 2005.
	
	\bibitem{brown2016metabolomics}
	Dustin~G Brown, Sangeeta Rao, Tiffany~L Weir, Joanne O’Malia, Marlon Bazan,
	Regina~J Brown, and Elizabeth~P Ryan.
	\newblock Metabolomics and metabolic pathway networks from human colorectal
	cancers, adjacent mucosa, and stool.
	\newblock {\em Cancer \& metabolism}, 4(1):11, 2016.
	
	\bibitem{Granovetter1973}
	Mark~S Granovetter.
	\newblock {The Strength of Weak Ties}.
	\newblock {\em American Journal of Sociology}, 78(6):1360--1380, 1973.
	
	\bibitem{Bassett2017}
	Danielle~S. Bassett and Olaf Sporns.
	\newblock {Network neuroscience}.
	\newblock {\em Nature Neuroscience}, 20(3):353--364, 2017.
	
	\bibitem{Barthelemy2011}
	Marc Barth{\'{e}}lemy.
	\newblock {Spatial networks}.
	\newblock {\em Physics Reports}, 499(1-3):1--101, 2011.
	
	\bibitem{Deville2016}
	Pierre Deville, Chaoming Song, Nathan Eagle, Vincent~D. Blondel,
	Albert-L{\'{a}}szl{\'{o}} Barab{\'{a}}si, and Dashun Wang.
	\newblock {Scaling identity connects human mobility and social interactions}.
	\newblock {\em Proceedings of the National Academy of Sciences},
	113(26):7047--7052, 2016.
	
	\bibitem{Sole2016}
	Albert Sol\'e-Ribalta, Sergio G\'omez, and Alex Arenas.
	\newblock Congestion induced by the structure of multiplex networks.
	\newblock {\em Phys. Rev. Lett.}, 116:108701, Mar 2016.
	
	\bibitem{Battiston2012}
	Stefano Battiston, Michelangelo Puliga, Rahul Kaushik, Paolo Tasca, and Guido
	Caldarelli.
	\newblock {DebtRank: Too Central to Fail? Financial Networks, the FED and
		Systemic Risk}.
	\newblock {\em Scientific Reports}, 2:541, aug 2012.
	
	\bibitem{Kefi2016a}
	Sonia K{\'{e}}fi, Vincent Miele, Evie~A. Wieters, Sergio~A. Navarrete, and
	Eric~L. Berlow.
	\newblock {How Structured Is the Entangled Bank? The Surprisingly Simple
		Organization of Multiplex Ecological Networks Leads to Increased Persistence
		and Resilience}.
	\newblock {\em PLoS Biology}, 14(8):1--21, 2016.
	
	\bibitem{Pilosof2017}
	Shai Pilosof, Mason~A Porter, Mercedes Pascual, and Sonia K{\'{e}}fi.
	\newblock {The multilayer nature of ecological networks}.
	\newblock {\em Nature Ecology and Evolution}, 1(4):1--9, 2017.
	
	\bibitem{arenas2008synchronization}
	Alex Arenas, Albert D{\'\i}az-Guilera, Jurgen Kurths, Yamir Moreno, and
	Changsong Zhou.
	\newblock Synchronization in complex networks.
	\newblock {\em Physics reports}, 469(3):93--153, 2008.
	
	\bibitem{Newman2002}
	Mark~E.J. Newman.
	\newblock {Spread of epidemic disease on networks}.
	\newblock {\em Phys. Rev. E}, 66(1):16128, 2002.
	
	\bibitem{Guardi02}
	X.~Guardiola, A.~D\'{\i}az-Guilera, C.~J. P\'erez, A.~Arenas, and M.~Llas.
	\newblock Modeling diffusion of innovations in a social network.
	\newblock {\em Phys. Rev. E}, 66:026121, Aug 2002.
	
	\bibitem{Barrat2008}
	Alain Barrat, Marc Barth{\'{e}}lemy, and Alessandro Vespignani.
	\newblock {\em {Dynamical Processes on Complex Networks}}.
	\newblock Cambridge University Press, 2008.
	
	\bibitem{PastorSatorras2015}
	Romualdo Pastor-Satorras, Claudio Castellano, Piet {Van Mieghem}, and
	Alessandro Vespignani.
	\newblock {Epidemic processes in complex networks}.
	\newblock {\em Reviews of Modern Physics}, 87(3):925--979, 2015.
	
	\bibitem{bottcher2017critical}
	Lucas B{\"o}ttcher, Jan Nagler, and Hans~J Herrmann.
	\newblock Critical behaviors in contagion dynamics.
	\newblock {\em Physical Review Letters}, 118(8):088301, 2017.
	
	\bibitem{jgg2018}
	Jesus G{\'o}mez-Garde{\~n}es, David Soriano-Pa{\~n}os, and Alex Arenas.
	\newblock Critical regimes driven by recurrent mobility patterns of
	reaction--diffusion processes in networks.
	\newblock {\em Nature Physics}, 14(4):391--395, 2018.
	
	\bibitem{Skardal19}
	Per~Sebastian Skardal and Alex Arenas.
	\newblock Abrupt desynchronization and extensive multistability in globally
	coupled oscillator simplexes.
	\newblock {\em Phys. Rev. Lett.}, 122:248301, Jun 2019.
	
	\bibitem{d2015anomalous}
	Raissa~M D’Souza and Jan Nagler.
	\newblock Anomalous critical and supercritical phenomena in explosive
	percolation.
	\newblock {\em Nature Physics}, 11(7):531, 2015.
	
	\bibitem{Hackett2016}
	A.~Hackett, D.~Cellai, S.~G\'omez, A.~Arenas, and J.~P. Gleeson.
	\newblock Bond percolation on multiplex networks.
	\newblock {\em Phys. Rev. X}, 6:021002, Apr 2016.
	
	\bibitem{Rapisardi2019}
	Giacomo Rapisardi, Alex Arenas, Guido Caldarelli, and Giulio Cimini.
	\newblock Fragility and anomalous susceptibility of weakly interacting
	networks.
	\newblock {\em Phys. Rev. E}, 99:042302, Apr 2019.
	
	\bibitem{molloy1995critical}
	Michael Molloy and Bruce Reed.
	\newblock A critical point for random graphs with a given degree sequence.
	\newblock {\em Random structures \& algorithms}, 6(2-3):161--180, 1995.
	
	\bibitem{DeDomenico2013}
	Manlio {De Domenico}, Albert Sol{\'{e}}-Ribalta, Emanuele Cozzo, Mikko
	Kivel{\"{a}}, Yamir Moreno, Mason~A Porter, Sergio G{\'{o}}mez, and Alex
	Arenas.
	\newblock {Mathematical Formulation of Multilayer Networks}.
	\newblock {\em Phys. Rev. X}, 3(4):41022, dec 2013.
	
	\bibitem{Perra2012}
	N.~Perra, B.~Gon{\c c}alves, R.~Pastor-Satorras, and A.~Vespignani.
	\newblock Activity driven modeling of time varying networks.
	\newblock {\em Scientific Reports}, 2:469 EP --, 06 2012.
	
	\bibitem{Valdano2018a}
	Eugenio Valdano, Michele~Re Fiorentin, Chiara Poletto, and Vittoria Colizza.
	\newblock {Epidemic Threshold in Continuous-Time Evolving Networks}.
	\newblock {\em Physical Review Letters}, 120(6):068302, 2018.
	
	\bibitem{Alessandretti2017}
	Laura Alessandretti, Kaiyuan Sun, Andrea Baronchelli, and Nicola Perra.
	\newblock {Random walks on activity-driven networks with attractiveness}.
	\newblock {\em Phys. Rev. E}, 95(5):52318, may 2017.
	
	\bibitem{Petri2018}
	Giovanni Petri and Alain Barrat.
	\newblock Simplicial activity driven model.
	\newblock {\em Phys. Rev. Lett.}, 121:228301, Nov 2018.
	
	\bibitem{newman2002assortative}
	Mark~E.J. Newman.
	\newblock Assortative mixing in networks.
	\newblock {\em Physical Review Letters}, 89(20):208701, 2002.
	
	\bibitem{Holland1983}
	Paul~W. Holland, Kathryn~Blackmond Laskey, and Samuel Leinhardt.
	\newblock {Stochastic blockmodels: First steps}.
	\newblock {\em Social Networks}, 5(2):109--137, jun 1983.
	
	\bibitem{Peixoto2017}
	Tiago~P Peixoto.
	\newblock {Bayesian stochastic blockmodeling}.
	\newblock In Ferligoj~A. {Doreian P., Batagelj V.}, editor, {\em Advances in
		Network Clustering and Blockmodeling}. Wiley, New York, NY, USA, 2018.
	
	\bibitem{lovasz1993random}
	L{\'a}szl{\'o} Lov{\'a}sz et~al.
	\newblock Random walks on graphs: A survey.
	\newblock {\em Combinatorics, Paul erdos is eighty}, 2(1):1--46, 1993.
	
	\bibitem{gomez2010discrete}
	Sergio G{\'o}mez, Alex Arenas, J~Borge-Holthoefer, Sandro Meloni, and Yamir
	Moreno.
	\newblock Discrete-time markov chain approach to contact-based disease
	spreading in complex networks.
	\newblock {\em EPL (Europhysics Letters)}, 89(3):38009, 2010.
	
	\bibitem{Granell13}
	Clara Granell, Sergio G\'omez, and Alex Arenas.
	\newblock Dynamical interplay between awareness and epidemic spreading in
	multiplex networks.
	\newblock {\em Phys. Rev. Lett.}, 111:128701, Sep 2013.
	
	\bibitem{Valdano2015b}
	Eugenio Valdano, Luca Ferreri, Chiara Poletto, and Vittoria Colizza.
	\newblock {Analytical Computation of the Epidemic Threshold on Temporal
		Networks}.
	\newblock {\em Phys. Rev. X}, 5(2):21005, apr 2015.
	
	\bibitem{PastorSatorras2001}
	Romualdo Pastor-Satorras and Alessandro Vespignani.
	\newblock {Epidemic spreading in scale-free networks.}
	\newblock {\em Phys. Rev. Lett.}, 86(14):3200--3203, 2001.
	
	\bibitem{Ferreri2014}
	Luca Ferreri, Paolo Bajardi, Mario Giacobini, Silvia Perazzo, and Ezio
	Venturino.
	\newblock {Interplay of network dynamics and heterogeneity of ties on spreading
		dynamics}.
	\newblock {\em Phys. Rev. E}, 90(1):12812, jul 2014.
	
	\bibitem{Dyson1949}
	F~J Dyson.
	\newblock {The Radiation theories of Tomonaga, Schwinger, and Feynman}.
	\newblock {\em Phys. Rev.}, 75:486--502, 1949.
	
	\bibitem{Blanes2009}
	S.~Blanes, F.~Casas, J.A. Oteo, and J.~Ros.
	\newblock {The Magnus expansion and some of its applications}.
	\newblock {\em Physics Reports}, 470(5-6):151--238, jan 2009.
	
	\bibitem{Argeri2014}
	Mario Argeri, Stefano {Di Vita}, Pierpaolo Mastrolia, Edoardo Mirabella,
	Johannes Schlenk, Ulrich Schubert, and Lorenzo Tancredi.
	\newblock {Magnus and Dyson series for master integrals}.
	\newblock {\em Proceedings of Science}, 2014-Janua, 2014.
	
	\bibitem{Tian2014}
	Jianjun~Paul Tian and Jin Wang.
	\newblock {Some results in Floquet theory, with application to periodic
		epidemic models}.
	\newblock {\em Applicable Analysis}, pages 1--25, 2014.
	
	\bibitem{mattila1999geometry}
	Pertti Mattila.
	\newblock {\em Geometry of sets and measures in Euclidean spaces: fractals and
		rectifiability}, volume~44.
	\newblock Cambridge university press, 1999.
	
	\bibitem{krioukov2010hyperbolic}
	Dmitri Krioukov, Fragkiskos Papadopoulos, Maksim Kitsak, Amin Vahdat, and
	Mari\'an Bogu\~n\'a.
	\newblock Hyperbolic geometry of complex networks.
	\newblock {\em Phys. Rev. E}, 82:036106, Sep 2010.
	
	\bibitem{Leskovec2005}
	Jure Leskovec, Deepayan Chakrabarti, Jon Kleinberg, and Christos Faloutsos.
	\newblock {Realistic, Mathematically Tractable Graph Generation and Evolution,
		Using Kronecker Multiplication}.
	\newblock In Al{\'{i}}pio~M{\'{a}}rio Jorge, Lu{\'{i}}s Torgo, Pavel Brazdil,
	Rui Camacho, and Jo{\~{a}}o Gama, editors, {\em Knowledge Discovery in
		Databases: PKDD 2005}, pages 133--145, Berlin, Heidelberg, 2005. Springer
	Berlin Heidelberg.
	
	\bibitem{Leskovec2007}
	Jure Leskovec and Christos Faloutsos.
	\newblock {Scalable Modeling of Real Graphs Using Kronecker Multiplication}.
	\newblock In {\em Proceedings of the 24th International Conference on Machine
		Learning}, ICML '07, pages 497--504, New York, NY, USA, 2007. ACM.
	
	\bibitem{Leskovec2010}
	Jure Leskovec, Deepayan Chakrabarti, Jon Kleinberg, Christos Faloutsos, and
	Zoubin Ghahramani.
	\newblock Kronecker graphs: An approach to modeling networks.
	\newblock {\em J. Mach. Learn. Res.}, 11:985--1042, March 2010.
	
	\bibitem{Kim2010}
	Myunghwan Kim and Jure Leskovec.
	\newblock {Multiplicative Attribute Graph Model of Real-World Networks}.
	\newblock In Ravi Kumar and Dandapani Sivakumar, editors, {\em Algorithms and
		Models for the Web-Graph}, pages 62--73, Berlin, Heidelberg, 2010. Springer
	Berlin Heidelberg.
	
	\bibitem{Seshadhri2013}
	Comandur Seshadhri, Ali Pinar, and Tamara~G Kolda.
	\newblock {An In-depth Analysis of Stochastic Kronecker Graphs}.
	\newblock {\em J. ACM}, 60(2):13:1----13:32, may 2013.
	
	\bibitem{Robles2016}
	Pablo Robles, Sebastian Moreno, and Jennifer Neville.
	\newblock Sampling of attributed networks from hierarchical generative models.
	\newblock In {\em Proceedings of the 22Nd ACM SIGKDD International Conference
		on Knowledge Discovery and Data Mining}, KDD '16, pages 1155--1164, New York,
	NY, USA, 2016. ACM.
	
	\bibitem{Mahdian2007}
	Mohammad Mahdian and Ying Xu.
	\newblock {Stochastic Kronecker Graphs}.
	\newblock In {\em Proceedings of the 5th International Conference on Algorithms
		and Models for the Web-graph}, WAW'07, pages 179--186, Berlin, Heidelberg,
	2007. Springer-Verlag.
	
	\bibitem{Estrada2008}
	Ernesto Estrada and Naomichi Hatano.
	\newblock {Communicability in complex networks}.
	\newblock {\em Phys. Rev. E}, 77(3):36111, mar 2008.
	
	\bibitem{Estrada2012}
	Ernesto Estrada, Naomichi Hatano, and Michele Benzi.
	\newblock {The physics of communicability in complex networks}.
	\newblock {\em Physics Reports}, 514(3):89--119, 2012.
	
	\bibitem{gauvin2014detecting}
	Laetitia Gauvin, Andr{\'e} Panisson, and Ciro Cattuto.
	\newblock Detecting the community structure and activity patterns of temporal
	networks: a non-negative tensor factorization approach.
	\newblock {\em PloS one}, 9(1):e86028, 2014.
	
	\bibitem{Arbenz1988b}
	Peter Arbenz and Gene~H Golub.
	\newblock {On the Spectral Decomposition of Hermitian Matrices Modified by Low
		Rank Perturbations with Applications}.
	\newblock {\em SIAM Journal on Matrix Analysis and Applications}, 9(1):40--58,
	1988.
	
	\bibitem{dynpas2001}
	Romualdo Pastor-Satorras, Alexei V\'azquez, and Alessandro Vespignani.
	\newblock Dynamical and correlation properties of the internet.
	\newblock {\em Phys. Rev. Lett.}, 87:258701, Nov 2001.
	
\end{thebibliography}


\end{document}